\documentclass[AMA,STIX1COL]{WileyNJD-v2}

\articletype{Article Type}

\received{}
\revised{}
\accepted{}

\usepackage{booktabs}
\usepackage[utf8]{inputenc}
\usepackage{hyperref}
\usepackage{url}
\usepackage{graphicx}
\usepackage{float}
\usepackage{amsmath}
\usepackage{multirow}
\usepackage[font=small,labelfont=bf]{subcaption}

\usepackage{amsfonts,dsfont}
\usepackage{multicol}
\usepackage{textcomp}
\usepackage{xcolor}
\usepackage{changepage}

\usepackage{algorithm}
\algnewcommand\algorithmicinput{\textbf{Input:}}
\algnewcommand\INPUT{\item[\algorithmicinput]}
\algnewcommand\algorithmicoutput{\textbf{Output:}}
\algnewcommand\OUTPUT{\item[\algorithmicoutput]}
\algdef{SE}[SUBALG]{Indent}{EndIndent}{}{\algorithmicend\ }

\makeatletter
\def\@fnsymbol#1{\ensuremath{\ifcase#1\or \dagger\or * \or \ddagger\or
   \mathsection\or \mathparagraph\or \|\or **\or \dagger\dagger
   \or \ddagger\ddagger \else\@ctrerr\fi}}
    \makeatother

\newcommand{\liu}[1]{{\color{black}{#1}}}
    
\begin{document}

\title{Distributed and Deep Vertical Federated Learning with Big Data}

\author[1]{Ji Liu}
\author[2]{Xuehai Zhou}
\author[3]{Lei Mo}
\author[4]{Shilei Ji}
\author[5]{Yuan Liao}
\author[6]{Zheng Li}
\author[7]{Qin Gu}
\author[8]{Dejing Dou}

\authormark{Liu \textsc{et al}}

\address[1]{\orgname{Baidu Inc.}, \orgaddress{\state{Beijing}, \country{China}}}
\address[3]{\orgname{School of Automation, Southeast University}, \orgaddress{\state{Nanjing}, \country{China}}}
\address[6]{\orgname{Research Center, Chengdu Medical Union Information Co.LTD}, \orgaddress{\state{Chengdu}, \country{China}}}
\address[7]{\orgname{Research Center, Chengdu Big Data Group Co.LTD}, \orgaddress{\state{Chengdu}, \country{China}}}

\corres{Ji Liu. \email{liuji04@baidu.com}}


\abstract[Summary]{
In recent years, data are typically distributed in multiple organizations while the data security is becoming increasingly important.
Federated Learning (FL), which enables multiple parties to collaboratively train a model without exchanging the raw data, has attracted more and more attention.
Based on the distribution of data, FL can be realized in three scenarios, i.e., horizontal, vertical, and hybrid.
In this paper, we propose to combine distributed machine learning techniques with Vertical FL and propose a Distributed Vertical Federated Learning (DVFL) approach.
The DVFL approach exploits a fully distributed architecture within each party in order to accelerate the training process.
In addition, we exploit Homomorphic Encryption (HE) to protect the data against honest-but-curious participants.
We conduct extensive experimentation in a large-scale cluster environment and a cloud environment in order to show the efficiency and scalability of our proposed approach. The experiments demonstrate the good scalability of our approach and the significant efficiency advantage (up to 6.8 times with a single server and \liu{15.1} times with multiple servers in terms of the training time) compared with \liu{baseline frameworks}. 
}


\maketitle

\section{Introduction}
Recent years have witnessed huge amounts of distributed data over Internet of Things (IoT) devices and organizations \cite{liu2022distributed}. However, data aggregation from distributed devices, multiple regions, or organizations, become almost impossible \cite{yang2019federated} due to the concerns of data security, stringent privacy and legal restrictions \cite{CCL, GDPR, chik2013singapore, CCPA, Gaff2014}.
In this case, training models with distributed data from various sources becomes critical for the collaboration among multiple parties. In order to address this issue, Federated Learning (FL) \cite{mcmahan2017communication, yang2019federated} is proposed. 

FL was first introduced to train a global model with unbalanced and non-Independent and Identically Distributed (non-IID) data distributed on mobile devices \cite{mcmahan2017communication}, which is referred to Horizontal Federated Learning (HFL). The concept of FL is extended to three categories, i.e., horizontal, vertical, hybrid \cite{yang2019federated}. 
The Vertical FL (VFL), which is also denoted as feature-based federated learning \cite{yang2019federated}, handles the decentralized data of the same identifications with different features while the hybrid FL deals with the data of different identifications and different features.
While many existing FL works focus on HFL,  VFL remains a critical open problem \cite{hardy2017private,nock2018entity}. VFL is well-fit for the cases with similar sample space but differs in feature space. For instance, an investment bank and a commercial bank in the same area may share the same group of clients (similar sample space), but the investment bank records the investment behaviors of users while the commercial bank stores consuming habits (difference in the feature space).

FL is an on-the-rising research area of privacy computing because jointly training a model in a privacy-preserving fashion is in high demand. In the field of privacy computing, the Secure Multi-Party Computation (MPC) \cite{Liu2020FedVision} and Trusted Execution Environment (TEE) \cite{Sabt2015TEE} are two mainstream research directions. The nature of TEE itself needs to provide an isolated runtime environment and secure storage for data, which compromises much of the computation efficiency. The overlapping of FL and MPC studies is ubiquitous, while MPC techniques like Homomorphic Encryption (HE), Oblivious Transfer (OT) or Private Set Intersection (PSI) are regularly transited to FL settings. Nevertheless, because of the limitation of MPC algorithms, even when they have been optimized, the computational speed is an order of magnitude slower than plaintext computing. Hence, shortening algorithm runtime and simultaneously ensuring data security become critical in industrial applications. In this case, a distributed deep learning structure for FL is becoming promising to address this issue. 

In this paper, we combine HE and PSI techniques from MPC studies to preserve data privacy. When the training data contains sensitive information, such as ages, facial features, health conditions, credit card numbers, etc., the direct usage of raw data becomes improper. As a widely used tool \cite{armknecht2015cryptoeprint, li2021security}, HE can improve the data security in FL settings. Since the data is preferably not be transmitted explicitly, researchers naturally suggest passing in the ciphered data. While it allows computation over encrypted data without access to the secret key, the HE technique has been generally utilized in the FL settings \cite{zhang2020batchcrypt}. Furthermore, before the training process, data from two parties need to be aligned using PSI. PSI methods allow two parties to obtain their data intersection without revealing additional information to another party \cite{pinkas2018scalable, chen2017fast}.

Recent years have witnessed the thriving of VFL studies in multiple domains, e.g., privacy-preserving methods, performance optimization, and application scenario extension, etc. VFL was first proposed with a Logistic Regression (LR) method in Taylor approximation model with partially HE \cite{hardy2017private}. 
A stochastic kernel learning model is also adapted to FL \cite{gu2020federated}. In addition, multiparty situation of VFL is studied in \cite{feng2020multi}. Some studies apply Gaussian Differential Private (DP) perturbation on LR or neural networks, which broadens privacy-preserving approaches on VFL \cite{chen2020vafl}. \liu{Vertical FL is exploited to collaboratively train a global model while ensuring the privacy and security of data. When multiple data owners conduct cross-silo federated learning, in order to protect the privacy and security of data, vertical federated learning generally utilizes cryptographic operations, which incur frequent encryption and decryption. While cross-silo network bandwidth is limited, the cryptographic operations bring significant overhead, and training time can be prolonged to several times to millions of times. Especially in industrial scenarios, such as advertising and marketing with the collaboration from multiple companies, it may take hours or even days to collaboratively train a global model with big data with cryptographic operations for security, which cannot meet the time requirements for rapid business development. The efficiency of vertical FL has become a critical problem that restricts its large-scale application \cite{guo2022microfedml,2021Kairouz}. Nonetheless, despite some of above approaches consider parallel execution, these approaches are not sufficient to handle industrial big data (generally over 10TB) in a deep learning setting. Although some frameworks, e.g., FATE \cite{FATE} and PyVertical \cite{romanini2021pyvertical}, enable the execution of VFL, they generally cannot exploit distributed resources, e.g., multiple workers, for efficient big data processing.}

\liu{In this paper, With the consideration of both efficiency and data privacy, we propose the Distributed Vertical Federated Learning (DVFL) approach, which exploits a distributed architecture to accelerate the training process of VFL with big data while combining HE to ensure data security. The proposed architecture can accelerate the training process of the vertical FL with an almost linear performance improvement, which can be applied in joint marketing advertising, data federation \cite{liu2021data} in government, joint risk control in financial companies and so on.}
\liu{We apply the parameter server and peer-to-peer communication architecture to accelerate the big data processing.}
The proposed architecture can be deployed in a large-scale cluster environment environment or a cloud environment \cite{de2019data}, where multiple powerful servers can be offered for parallel computing \cite{liu2022large}.

We summarise our main contributions of this paper as follows:
\begin{itemize}
\item A fully Distributed structure for Vertical Federated Learning (DVFL) in order to handle big data in industrial applications.

\item The combination of the DVFL and HE techniques to protect the security and privacy of data.

\item We carry out extensive experimentation based on real-life deep learning models in a distributed environment, including the large-scale cluster environment and the cloud environment, which shows the advantages of DVFL over state-of-the-art platforms, i.e., FATE and PyVertical.
\end{itemize}

The rest of this paper is organized as follows. In Section \ref{sec:relatedwork}, we give a brief overview of related works. In section \ref{sec:preliminaries}, we list the preliminaries. 
In section \ref{sec:solution}, we present our distributed solution, i.e., DVFL. In section \ref{sec:experiment}, we present the experiment details and results. In section \ref{sec:conclusion}, we draw our conclusion and shed light on possible future research work.

\section{Related Work}
\label{sec:relatedwork}
In 2016, the concept of FL was first introduced by \cite{mcmahan2017communication}. Their approach aims at training a Logistic Regression (LR) model based on distributed data stored in a huge amount of global mobile users without conflicting local regulations or laws. Their approach is classified into horizontal FL years later, because of the wide range of participants and similar data features\liu{, e.g., health care with the attention mechanism on graphs \cite{ahmed2022hyper}, 5G-empowered drone networks with reinforcement learning for smart grid or smart cities \cite{ahmed20215g}, frequent itemset mining \cite{ahmed2021federated}, distributed medical data \cite{gaurav2022security} and education data \cite{rajabi2019exposing}, purchase behaviour with an attention mechanism \cite{ahmed2022reliable}, and Internet of Things (IoT) devices \cite{sejdiu2020integration,gaurav2022security}. In addition, the horizontal FL can be carried out in the Cloud for extreme gradient boosting \cite{wang2020cloud} or deep learning models with the combination of synchronous \cite{liu2022Efficient,liu2022multi,Zhang2022FedDUAP,che2022federated,jin2022accelerated} and asynchronous mechanisms \cite{Li2022FedHiSyn,stergiou2021infemo}.} Later, the first FL on vertically partitioned data was proposed to train a LR model \cite{hardy2017private}, which exploits HE to protect data security. Lastly, A GELU-Net model is trained based on the combination of FL and HE, while the experimentation was carried out using the MNIST dataset, which shows a decent computation speed performance \cite{zhang2018gelu}. In addition, the 128 MB MNIST dataset (MNIST) is far smaller than industrial big data. Hitherto, upon state-of-the-art VFL models, there is not a solution that can efficiently handle real big data with the VFL settings.

We have witnessed the growth of the distributed learning over past years. Distributed models like DistBelief \cite{dean2012large}, GraphLab \cite{low2014graphlab}, Petuum \cite{xing2015petuum}, Naiad \cite{murray2013naiad}, MLbase \cite{kraska2013mlbase}, etc, have been well explored in the early years. A parameter server architecture \cite{liu2021heterps} can utilize a parameter server or several parameter servers and multiple workers to train a model using a dataset of 636 TB in around 1 hour \cite{li2013parameter,li2014communication}. The parameter servers coordinates the execution while workers work in parallel to train the model. We were inspired by their parameter server architecture to propose the DVFL. 

There has been a fairly long history of PSI study \cite{freeman2004privatematching}. PSI has been widely used as the entity resolution protocol in VFL researches \cite{lu2020multi}. An efficient PSI protocol that combines Bloom Filter (see details in Section \ref{psi}) with secret share scheme \cite{dong2013private} is used in DVFL. Using the PSI techniques, we can transform a hybrid FL problem to a VFL problem while ensuring the data security.

Lastly, HE has been widely used in VFL settings. Two main branches of HE exist, i.e., fully HE and partially HE. The fully HE, e.g., Brakerski-Gentry-Vaikuntanathan HE \cite{gentry2009fully}, supports both addition and multiplication on ciphertext, while partially HE, e.g., Paillier HE \cite{paillier1999public}, supports either addition or multiplication operation on ciphertext. Both the fully and partially HE can be exploited with the VFL settings while the fully HE has the problem of high time complexity \cite{hardy2017private}. While it has better runtime efficiency, partially HE corresponds to less computational flexibility. The property of the Paillier cryptosystem allows addition on both parties of ciphertext and multiplication on one party. Paillier HE is widely utilized as the privacy-preserving strategy \cite{zhang2018gelu}. 

\section{Preliminaries}\label{sec:preliminaries}
In this section, we present the background of DVFL. We first introduce the concept of PSI and the idea of its protocol. Then, we explicate the detail of the parameter server. Afterward, we indicate Paillier HE and its homomorphic property. Finally, we specify the deep neural networks structure and how we apply the HE technique in DVFL. 

\subsection{Private Set Intersection}
\label{psi}

\begin{figure}[t]
    \centering
    \includegraphics[width=7cm]{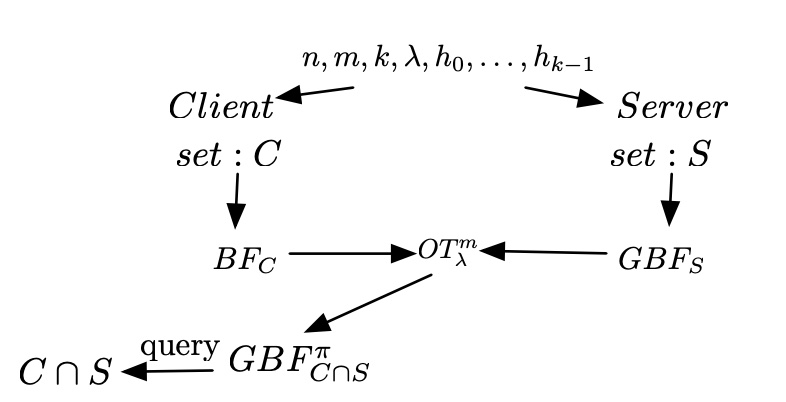}
    \caption{The basic PSI protocol \cite{dong2013private}}
    \label{fig:psi}
\end{figure}

Private Set Intersection (PSI) refers to a solution that calculates the intersection of two parties. Each of the two parties holds a set of items, and the two parties want to know the intersection of their sets without revealing any information outside of the intersection. Let us assume that one party has a set of items A = \{$a_1$, $a_2$, ..., $a_n$\} and another has a set of B = \{$b_1$, $b_2$, ..., $b_m$\}. PSI is a protocol that acquires the intersection of the two parties, i.e., A $\cap$ B, while 
it does not share any data excluded from the intersection of the two sets. If $a_i$ $\notin$ B, then party B should learn nothing about $a_i$ with a PSI solution. 

As shown in Figure \ref{fig:psi}, the PSI algorithm \cite{dong2013private} we exploit is constructed using two Bloom Filters (BF) where the BF is designed to tell whether an element is present in a set \cite{broder2004network}. Let us assume that one party is a server with set S, and another party is a client with set C. The client computes a BF that encodes its set C and the server computes a Garbled Bloom Filter (GBF) that encodes its set S. 
Note that the GBF is conceptually a BF with a security parameter while GBF has better capacity to hide the information of the items in the union of the two sets but outside of the intersection of the two sets. Once the GBF and the BF are obtained, the server acts as a sender and the client acts as a receiver to execute an OT protocol \cite{naor2001efficient}. 
The OT protocol allows a receiver to obtain one out of two strings held by a sender, without revealing to the sender the identity of its selection \cite{ishai2008founding}. Then, the server and the client can build the intersection $GBF_{C \cap S}$ on $GBF_S$ and $BF_C$. Thus, the client computes the ${C \cap S}$ set by querying all elements in its set filtered by the $GBF_{C \cap S}$. BF has an intrinsic flaw of false positive cases, while the false positive is negligible \cite{dong2013private}. In addition, since the data is independently hashed, it is a good single program multiple data (SPMD) scenario to run this algorithm in parallel.

\subsection{Parameter Server}
\label{parameterServer}

\begin{figure}[t]
\centering
\includegraphics[width=7cm]{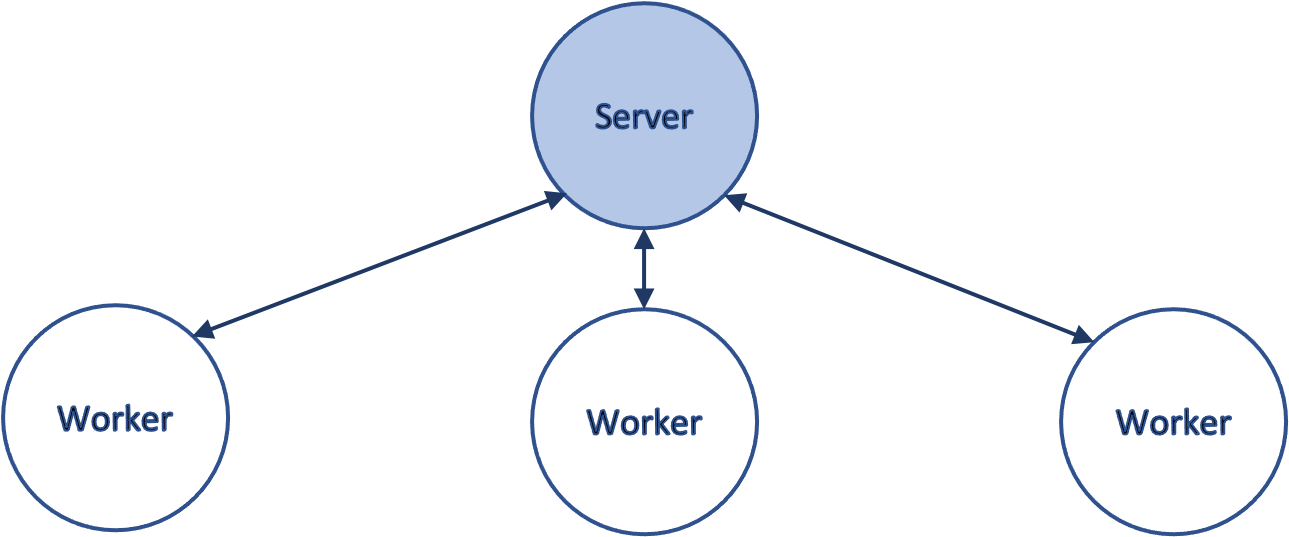}
\caption{Parameter Server Architecture}
\label{fig:ps}
\end{figure}

The parameter server architecture \cite{smola2010architecture} consists of a server and a group of workers, which moves data processing from one computing node to multiple computing nodes. Partitioned data are passed into paralleled workers. Each worker does not mutually communicate with other workers, but only communicates with the server node to retrieve and update gradients. Workers can run on GPUs or CPUs while the workers we utilized are CPU-oriented. Figure \ref{fig:ps} gives the intuition of the parameter server architecture while the number of workers could increase or decrease in different application scenarios. 

When there are many layers in a neural networks, the gradients are divided into multiple chunks to be communicated. These chunks of gradients need to be indexed, and the shared parameters are stored as key-value pairs. The key represents the ID, and the value represents the weight. Two operations are defined to bridge the server-worker communication: 
\begin{itemize}
\item \emph{push}: a worker sends its computed weights/gradients to the server.
\item \emph{pull}: a worker retrieves the aggregated weights/gradients from the server.
\end{itemize}

The distributed weights/gradients in workers are pushed and aggregated in the server, e.g., by calculating the average of the pushed weights or gradients. Then workers pull the combined gradient from the server. The server and the workers run these two operations until the model converges.

\subsection{Paillier Cryptosystem}
\label{paillier}
HE allows data operates computation, sort, search, edit on the ciphertext. As mentioned earlier, fully HE and partially HE are two HE main branches. Fully HE supports multiple operations of unbounded depth on ciphertext, which is the strongest notion of HE. However, it is unpractical to run fully HE on big data because of its high time complexity. Even through partially HE supports only one operation (addition or multiplication, not both) on ciphertext, researchers tend to use partially HE to achieve a better runtime performance. 

Paillier cryptosystem is a well-known partially HE, which is invented by and named after Pascal Paillier \cite{paillier1999public}. Let us define plaintext numbers $m_1$ and $m_2$ and choose two large prime \emph{p} and \emph{q}. Then let \emph{n} to be the multiplication of \emph{p} and \emph{q}. Paillier's homomorphic property allows addition of two ciphertexts and multiplication of a ciphertext by a plaintext. When two ciphertexts are multiplied, the result decrypts to the sum of their plaintext as shown in Equation \ref{eqn:homoAdd}. 
\begin{equation} \label{eqn:homoAdd}
    D_{priv}(E_{pub}(m_1) \times E_{pub}(m_2) \mod n^2) = m_1 + m_2 \mod n,
\end{equation}
where $D_{priv}$ represents the decryption operation, and $E_{pub}$ represents the encryption operation. 
When a ciphertext is raised to the power of plaintext, the result decrypts to the product of the two plaintexts as shown in Equation \ref{eqn:homoMulti}.
\begin{equation} \label{eqn:homoMulti}
D_{priv}(E_{pub}(m_1)^{m_2} \mod n^2) = m_1 \times m_2 \mod n
\end{equation}

\subsection{Deep Neural Networks}
\label{dnn}

\begin{figure*}[t]
    \centering
    \includegraphics[width=0.7\textwidth]{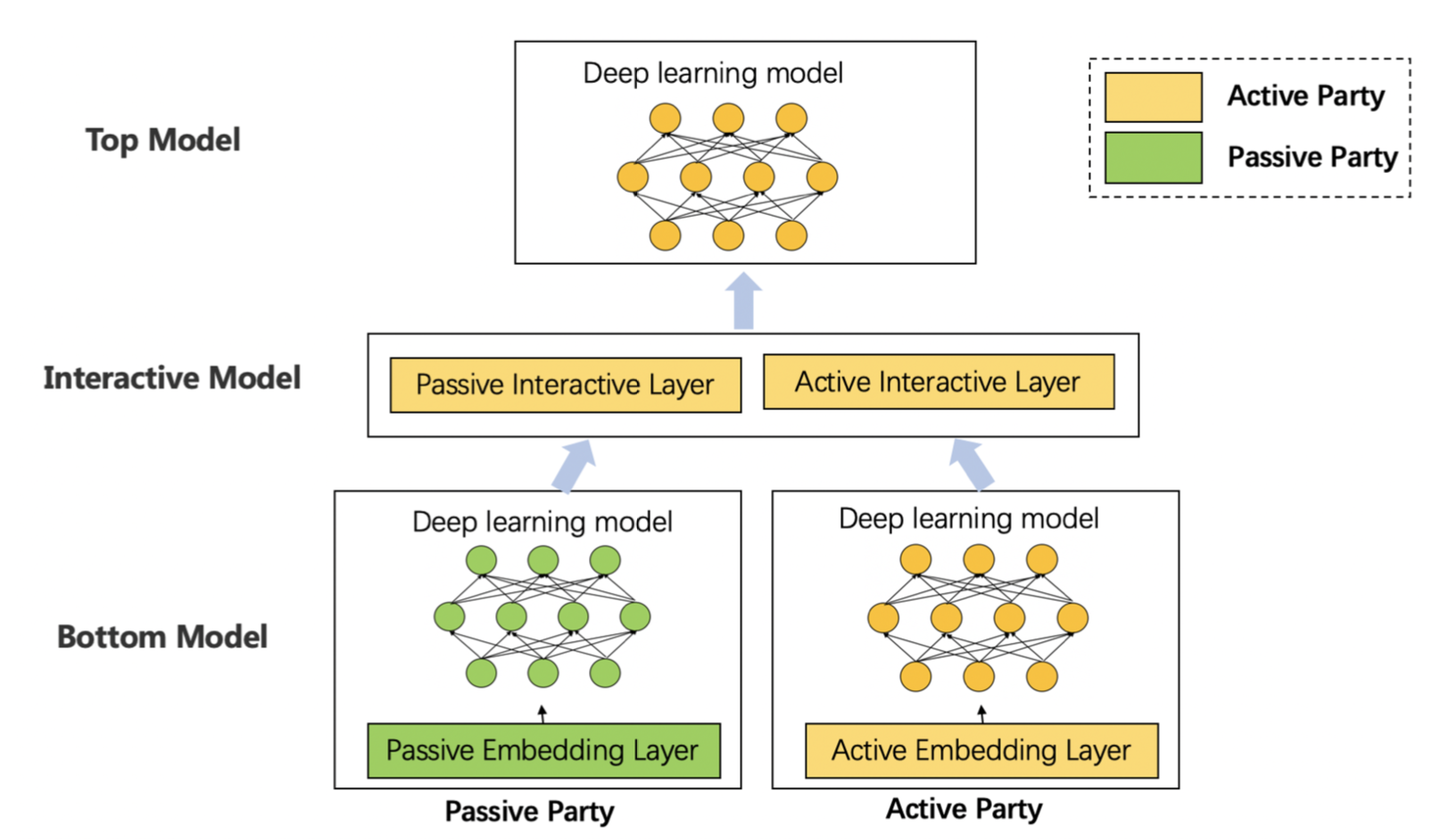}
    \caption{Deep Neural Networks for vertical FL \cite{zhang2018gelu}}
    \label{fig:dnnModel}
\end{figure*}

The deep learning structure shown in Figure \ref{fig:dnnModel} is constructed by three levels: bottom neural networks, interactive layer, and top neural networks \cite{zhang2018gelu}. A bottom neural network is given to any party that provides feature data. It brings more scalability to the architecture when more parties may participate in. Note that the bottom model can be a neural network with an arbitrary number of layers and an arbitrary number of neurons per layer. Then bottom networks are joined by a fully connected interactive layer. 

Let the party that holds the label matrix be the active party, while the party that only holds a feature matrix be the passive party. Inside the interactive layer, both passive and active parties generate noise and encrypt their data by HE. Then the merged output are passed to the top model. The top model calculates the loss of the model and propagates the error back. The merit of this structure is that all cross-institutional data communications are processed within the interactive layer, which significantly reduces the risk of data leakage.

\section{DVFL Solution}
\label{sec:solution}

In this section, we illuminate how the approaches mentioned in Section \ref{sec:preliminaries} are combined in DVFL. Moreover, we propose our solution, i.e., Distributed Vertical Federated Learning (DVFL). 

We propose DVFL to solve the problem of excessive runtime given by existing models on vertically partitioned data when it comes to big data. We exploit the distributed architecture to address this issue. We exploit the parameter server architecture explained in Section \ref{parameterServer} for the parallelism. 

\begin{figure*}[ht]
    \centering
    \includegraphics[width=0.75\textwidth]{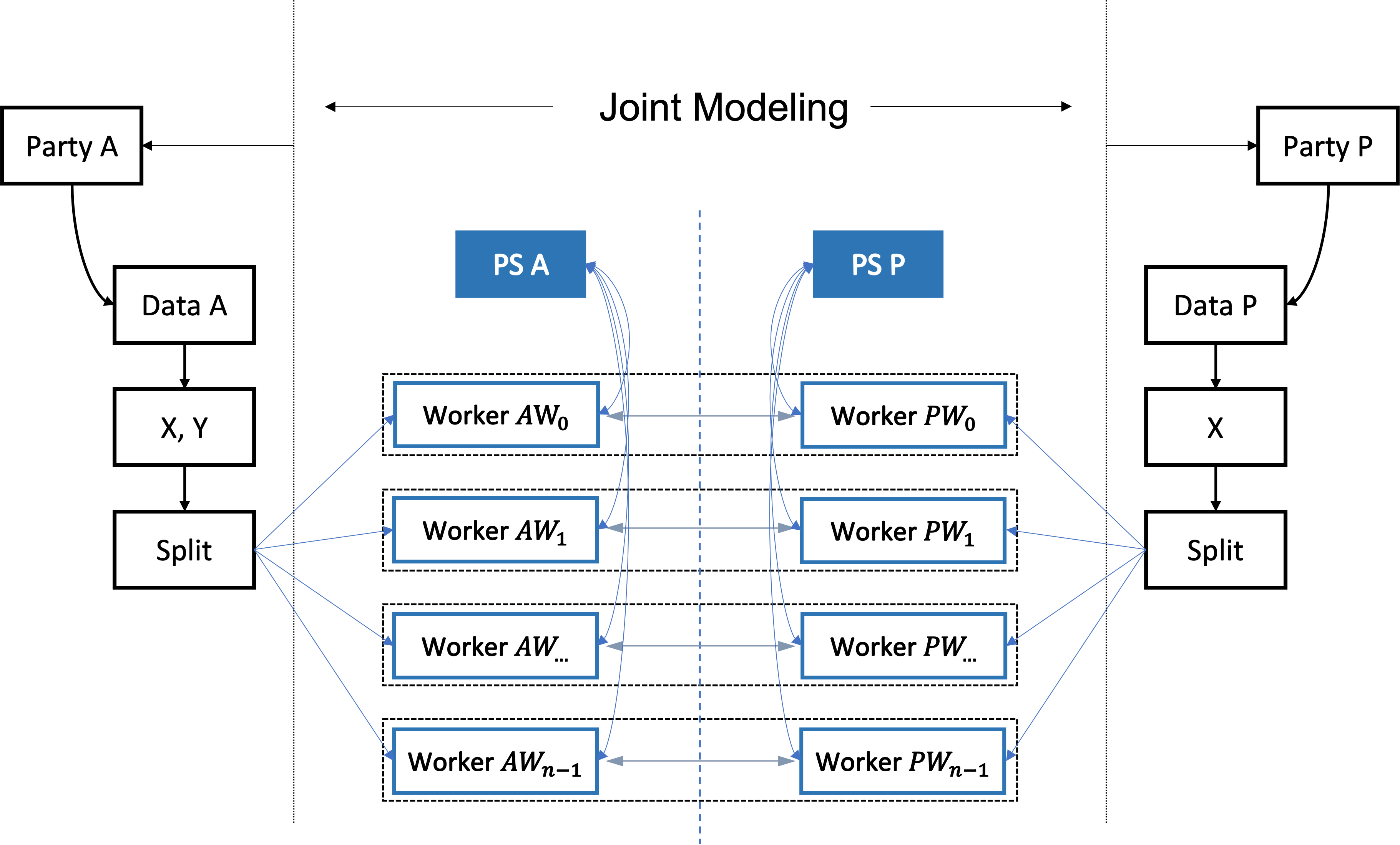}
    \caption{Distributed architecture with parameter servers for DVFL.}
    \label{fig:distributed}
\end{figure*}

\begin{algorithm}[t]
\caption{DVFL}
\label{alg:dvfl}
\begin{algorithmic}[1]
\INPUT ~\newline
\emph{$(X_A, Y_A)$}: active dataset; \newline
\emph{$X'_P$}: passive dataset; \newline
$n$: the number of workers in a party
\OUTPUT ~\newline
$Net$: a trained network

\State \emph{$(X_{A \cap P} , Y_{A \cap P})$}, \emph{$X'_{A \cap P}$} $\leftarrow$ DistributedPSI(\emph{$(X_A, Y_A)$}, \emph{$X'_P$})\label{alg:psi}
\State \{($X_1$, $Y_1$), ($X_2$, $Y_2$), ..., ($X_n$, $Y_n$)\}, \{$X_1^{\prime}$,$X_2^{\prime}$, ...,$X_n^{\prime}$\} $\leftarrow$ sequentialPartition(\emph{$(X_{A \cap P} , Y_{A \cap P})$}, \emph{$X'_{A \cap P}$}) 
\Comment{partition the data into $n$ similar length subsets} \label{alg:spartition}
\State $Net \leftarrow$ VerticalFL(\{($X_1$, $Y_1$), ($X_2$, $Y_2$), ..., ($X_n$, $Y_n$)\}, \{$X_1^{\prime}$,$X_2^{\prime}$, ...,$X_n^{\prime}$\}) \label{alg:vflInDVFL}
\end{algorithmic}
\end{algorithm}
We adopt the deep learning architecture discussed in Section \ref{dnn}. DVFL supports joint training with two parties, i.e., an active party and a passive party. The active party holds the feature matrix and the label matrix, and dominates the computation in training. The passive party holds only the feature matrix. We denote the active party Party A and the passive party Party P. Figure \ref{fig:distributed} shows the distributed architecture of DVFL. Data from Party A and Party P are partitioned to the number of n workers, \{($X_1$, $Y_1$), ($X_2$, $Y_2$), ..., ($X_n$, $Y_n$)\} and \{$X_1^{\prime}$,$X_2^{\prime}$, ...,$X_n^{\prime}$\}. A parameter server is applied to each party. Workers under the same server do not communicate with each others. But they communicate with their corresponding workers across parties with a peer-to-peer manner for the PSI process and distributed training process of vertical FL.

\begin{algorithm}[t]
\caption{Distributed PSI}
\label{alg:distributedPSI}
\begin{algorithmic}[1]
\INPUT ~\newline
\emph{$A$}: identification set in the active dataset; \newline
\emph{$P$}: identification set in the passive dataset; \newline
$n$: the number of workers in a party
\OUTPUT ~\newline
$A$ $\cap$ $P$: the intersection of $A$ and $P$
\State \{$A_1$, $A_2$, ..., $A_n$\} $\leftarrow$ hashPartition(\emph{$A$}) 
\Comment{Partition the identification set in the active party into $n$ similar length subsets using a hash function} \label{alg:activeHashPartition}
\State \{$P_1$, $P_2$, ..., $P_n$\} $\leftarrow$ hashPartition(\emph{$P$}) 
\Comment{Partition the identification set in the passive party into $n$ similar length subsets using the same hash function} \label{alg:passiveHashPartition}
\For{$i \in$ \{1, 2, ..., n\}}\label{alg:forBegin} \Comment{execution in parallel}
\State $A_i$ $\cap$ $P_i$ $\leftarrow$ PSI(\{$A_i$, $P_i$\}) \label{alg:PSI}
\EndFor\label{alg:forEnd}
\State $A$ $\cap$ $P$ $\leftarrow$ ($A_1$ $\cap$ $P_1$) $\cup$ ($A_2$ $\cap$ $P_2$) $\cup$ ...  $\cup$ ($A_n$ $\cap$ $P_n$)\label{alg:unionResults}
\end{algorithmic}
\end{algorithm}

The procedure of DVFL is demonstrated in Algorithm \ref{alg:dvfl}.
$(X_A, Y_A)$ represents the active dataset with $X_A$ representing the features of the data set with identification set ($A$) in the sample space of the active party and $Y_A$ representing the corresponding label for each data item. 
$X'_P$ represents the passive dataset with identification set $P$ in the sample space of the passive party. 
Once the dataset components are clarified, there are two preparation jobs before the training process of vertical FL. The first job is that two parties securely apply a Distributed PSI protocol to their data in order to obtain the common sample space ($A \cap P$) (Line \ref{alg:psi}, see details in Algorithm \ref{alg:distributedPSI}). The second job is that two parties utilize a sequential data partitioning strategy to partition the big data into small chunks and send them to the workers of the parameter server architecture (Line \ref{alg:spartition}). The items in each pair of data chunk with the same index in the active party and the passive party share the same identifications. Finally, a neural network is trained using the distributed vertical FL setting (Line \ref{alg:vflInDVFL}) as explained in Algorithm \ref{alg:VFLA}.

\begin{algorithm}[t]
\caption{Vertical FL}
\label{alg:VFLA}
\begin{algorithmic}[1]
\INPUT ~\newline
\{($X_1$, $Y_1$), ($X_2$, $Y_2$), ..., ($X_n$, $Y_n$)\}: dataset in the active party; \newline
\{$X_1^{\prime}$,$X_2^{\prime}$, ...,$X_n^{\prime}$\}: dataset in the passive party; \newline
$Net_o$: an initial neural network

\OUTPUT ~\newline
$Net$: a trained neural network

\State $Net \leftarrow Net_o$\label{alg:netInit}
\While{Stop condition is not met}\label{alg:whileLoop}
\For{$i \in$ \{1, 2, ..., n\}} \Comment{execution in parallel}\label{alg:forLoop}
\State $w_i$ $\leftarrow$ getWeight($X_i$, $Y_i$, $X'_i$)\label{alg:getWeight}
\EndFor
\State $Net$ $\leftarrow$ update($Net$, $\{w_1, w_2, ..., w_n\}$)
\EndWhile
\end{algorithmic}
\end{algorithm}

Algorithm \ref{alg:distributedPSI} illustrates the distributed PSI. First, the identification sets of active party (Line \ref{alg:activeHashPartition}) and passive party (Line \ref{alg:passiveHashPartition}) are partitioned to $n$ buckets of identifications with $n$ representing the number of workers in each party using the same hash function \cite{shasha1991optimizing}.
Then, each bucket of identifications is handed by a corresponding worker, e.g., $A_i$ is handled by Work $i$ in the active party and $P_i$ is handled by Work $i$ in the passive party.
As the identifications are partitioned using the same hash function, the same identifications in active dataset and passive dataset should be partitioned into the buckets of the same index.
Then, the PSI protocol can be performed between the corresponding workers of the same index in active and passive parties. Then, the PSI protocol is realized within each pair of workers in parallel, i.e., the function $PSI$ (Line \ref{alg:psi}) in the loop (Lines \ref{alg:forBegin} - \ref{alg:forEnd}). Finally, the intersection between $A$ and $P$ is obtained as the union of the intersection results of each pair of identification buckets within the parameter servers in the active and passive parties.

\begin{algorithm}[t]
\caption{Distributed Training in the Active Party}
\label{alg:activeDistributedVFL}
\begin{algorithmic}[1]
\INPUT ~\newline
$PW_i$: Worker $i$ in the passive party; \newline
$X_i$: the features of dataset $i$ in $AW_i$; \newline
$Y_i$: the labels of dataset $i$ in $AW_i$
\State $aw_{global}$ $\leftarrow$ $\emph{pull()}$ \label{alg:pullActive}
\State $activeResult$ $\leftarrow$ forwardPropagation($aw_{global}$, $X_i$, activeBottomModel) \label{alg:forward}
\State $securedActiveResult$ $\leftarrow$ HE($activeResult$) \label{alg:secureActive}
\State $passiveResult$ $\leftarrow$ receiveResult($PW_i$) \label{alg:receive}
\State $result$ $\leftarrow$ forwardPropagation($aw_{global}$, $securedActiveResult$, $passiveResult$, $activeTopModel$) \label{alg:forwardTop}
\State $activeTopLoss, activeTopGradient$ $\leftarrow$ backPropagation($result$, $activeTopModel$, $aw_{global}$, $Y_i$) \label{alg:backTop}
\State sendLoss($activeTopLoss$, $PW_i$) \label{alg:send}
\State $activeBottomGradient$ $\leftarrow$ backPropagation($activeTopLoss$, $activeBottomModel$, $aw_{global}$) \label{alg:backBottom}
\State $updatedWeights$ $\leftarrow$ gradientDescent($activeTopGradient$, $activeBottomGradient$, $ag_{global}$)  \label{alg:gradient}
\State sendWeightsToActivePS($updatedWeights$)
\end{algorithmic}
\end{algorithm}

\begin{algorithm}[t]
\caption{Distributed Training in the Passive Party}
\label{alg:passiveDistributedVFL}
\begin{algorithmic}[1]
\INPUT ~\newline
$AW_i$: Worker $i$ in the active party; \newline
$X'_i$: the features of dataset $i$ in $PW_i$
\State $pw_{global}$ $\leftarrow$ $\emph{pull()}$ \label{alg:pullPassive}
\State $activeResult$ $\leftarrow$ forwardPropagation($pw_{global}$, $X'_i$, passiveBottomModel) \label{alg:forwardPassive}
\State $securedActiveResult$ $\leftarrow$ HE($activeResult$) \label{alg:securePassive}
\State sendResult($AW_i$, $securedActiveResult$) \label{alg:sendPassive}
\State $pg$ $\leftarrow$ receiveLoss($AW_i$) \label{alg:receivePassive}
\State $passiveBottomGradient$ $\leftarrow$ backPropagation($pg$, $passiveBottomModel$, $pw_{global}$) \label{alg:backBottomPassive}
\State $updatedWeights$ $\leftarrow$ gradientDescent( $passiveBottomGradient$, $pg_{global}$)  \label{alg:gradientPassive}
\State sendWeightsToPassivePS($updatedWeights$)
\end{algorithmic}
\end{algorithm}

The training process of the vertical FL is shown in Algorithm \ref{alg:VFLA}. First, the neural network (model) is randomly initialized (Line \ref{alg:netInit}). Second, when the stop condition, e.g., required accuracy or a predefined number of iterations, is not met (Line \ref{alg:whileLoop}), e.g., more iterations should be executed, the model continues to be updated. Then, each pair of workers in the active and passive parties calculates the weights of the model (Lines \ref{alg:forLoop} and \ref{alg:getWeight}), which can be realized in parallel. Afterward, the workers send the weights to the parameter servers in each party, and the parameter servers update the model using a model aggregation method, such as BSP \cite{zhao2019elastic}. During the training process, each worker computes a local gradient on its given data, and the server aggregate these gradients to update the global parameters in each iteration. Workers pull the updated global parameter from the server and start the next round of iteration. The parameter server executes a great number of gradient exchanges to update the parameter until the model converges. The details of distributed training at each worker of the active or passive party are explained in Algorithms \ref{alg:activeDistributedVFL} and \ref{alg:passiveDistributedVFL}, respectively.

The training process of the distributed vertical FL in the active party is illustrated in Algorithms \ref{alg:activeDistributedVFL}.
First, the global weights are pulled from the parameter server in the active party (Line \ref{alg:pullActive}). 
Then, forward propagation is carried out with the bottom model and the intermediate result is generated (Line \ref{alg:forward}).  
Afterward, after receiving the result from the worker in the passive party (Line \ref{alg:receive}), and encrypting the intermediate result using the paillier cryptosystem as explained in Section \ref{paillier} (Line \ref{alg:secureActive}), the forward propagation is carried out with the top model (Line \ref{alg:forwardTop}) and the result of the whole model is produced.
We exploit the encryption and decryption detailed in \cite{zhang2018gelu} for the interaction between any pair of active worker and passive worker.
The loss and gradients can be calculated using the result and the labels in the active party with the top model (Line \ref{alg:backTop}).
The intermediate loss is sent to the worker in the passive party (Line \ref{alg:send}).
In addition, the gradient for the bottom model can be generated through the back propagation of the bottom model (Line \ref{alg:backBottom}). 
The model is updated with the gradient using gradient descent (Line \ref{alg:gradient}).
Finally, the updated weights are sent to the parameter server in the active party.

The training process of the distributed vertical FL in the passive party is shown in Algorithms \ref{alg:passiveDistributedVFL}.
First, the global weights are pulled from the parameter server in the active party (Line \ref{alg:pullPassive}). 
Then, forward propagation is carried out with the bottom model and the intermediate result is generated (Line \ref{alg:forwardPassive}).  
The result is encrypted using the paillier cryptosystem as explained in Section \ref{paillier} (Line \ref{alg:securePassive}) before being sent to the counterparty worker in the active party (Line \ref{alg:sendPassive}).
When receiving the gradients from the worker in the active party (Line \ref{alg:sendPassive}), 
Afterward, after receiving the loss from the worker in the passive party (Line \ref{alg:receivePassive}), backward propagation is carried out to calculate the gradients for the bottom model in the passive party (Line \ref{alg:backBottomPassive}).
The model is updated with the generated gradients (Line \ref{alg:gradientPassive}), and the updated weights are sent to the parameter server in the passive party. 

\liu{As we exploit the hash function to split the data, the overhead of the data split is $\mathcal{O}(1)$. In addition, in Line 6 of Algorithm \ref{alg:VFLA}, the models should be aggregated for the model update, thus, the overhead to calculate the averaged model is $\mathcal{O}(n)$ with $n$ representing the number of cores in each party. Finally, the overall complexity of DVFL is $\mathcal{O}(n)$.}

\section{Experiment}
\label{sec:experiment}

In this section, we present the results of experiments in a large-scale cluster environment and the Baidu AI cloud environment. We illustrate our data source as well as data features and labels. We account for how much impact to run not only the training process of vertical FL but also the PSI protocol parallel with different number of workers. Then, we discuss the bottleneck of the experiment between CPU and memory usage under Baidu AI Cloud environment. Afterward, we compare the efficiency of DVFL and state-of-the-art platforms, i.e., FATE \cite{FATE} and PyVertical \cite{romanini2021pyvertical}, with different amounts of data and numbers of workers on the Baidu AI cloud environment. We fine tuned the parameters for both FATE and DVFL.

We implemented a prototype based on the architecture of DVFL. The data communication in DVFL between any two of the parameter servers and workers is implemented by a Remote Procedure Call (RPC), i.e., gRPC \cite{GRPC}, and the data communication is realized using Socket \cite{kalita2014socket}. In the following figures, we denote $n$ parties with $m$ workers at each party by $n * m$. We exploit the ''a9a'' dataset from the LIBSVM library \cite{chang2011libsvm} in the experimentation.

\subsection{DVFL}
\label{dvfl}

\begin{figure}[t] 
    \centering
    \subcaptionbox{The training time of the distributed VFL with different numbers of workers.}{\includegraphics[width=0.45\textwidth]{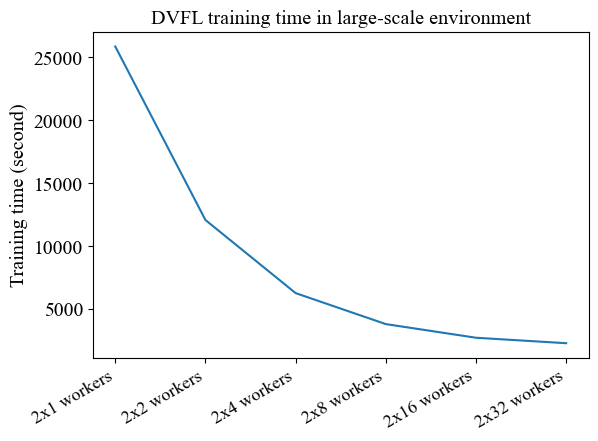}}
    \hfill 
    \subcaptionbox{The throughput of the distributed VFL with different numbers of workers.}{\includegraphics[width=0.45\textwidth]{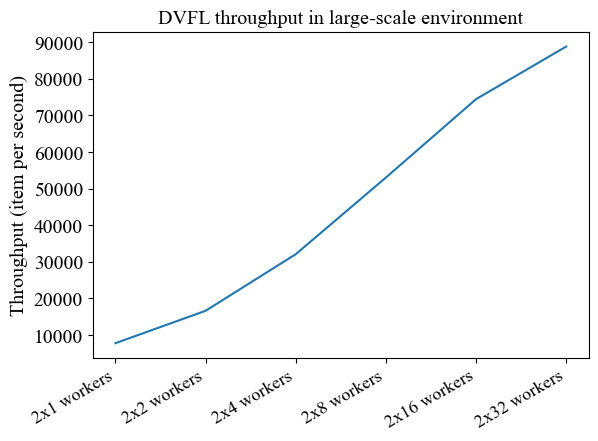}}
    \caption{The performance of distributed VFL in the large-scale cluster environment.
    As the number of worker nodes increases, the execution time of the distributed VFL is reduce, and the throughput is improved.}
    \label{performanceDVFL}
\end{figure}
The large-scale cluster environment is given by Hygon C86 7185 thirty-two 6000-core processors, 128 GB memory, four 6000-core Distributed Control Unit (DCU), 64-bit CentOS 7.6 Linux with x86\_64 architecture. To run DVFL, Party A and Party P each contains $10^6$ rows of data. To evaluate the system performance, we record and compare the time consumption of two parties running DVFL under 1, 2, 4, 8, 16, 32 workers respectively. As it shown in Figure \ref{performanceDVFL}, with the number of workers increase, the overall runtime drops while the total data processed per second increases. The execution time given by non-distributed system (1-worker per party) is 25865 seconds, while the execution time given by 32-worker per party is 2252 seconds. And the data processed per second is increased from 7732 rows (1-worker) to 88810 (32-worker). The assumption of the viability of DVFL is confirmed by noticing the tremendous execution time drop and the immense data processing capacity by the multi-worker system. 

\begin{figure}[t] 
    \centering
    \subcaptionbox{The execution time of distributed PSI with different numbers of workers.}{\includegraphics[width=0.45\textwidth]{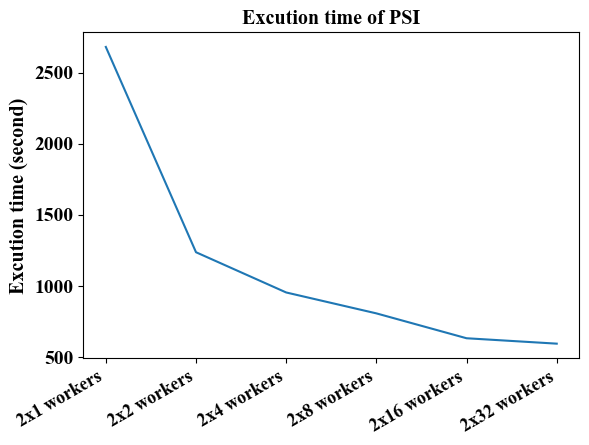}}
    \hfill 
    \subcaptionbox{The throughput of distributed PSI with different numbers of workers.}{\includegraphics[width=0.45\textwidth]{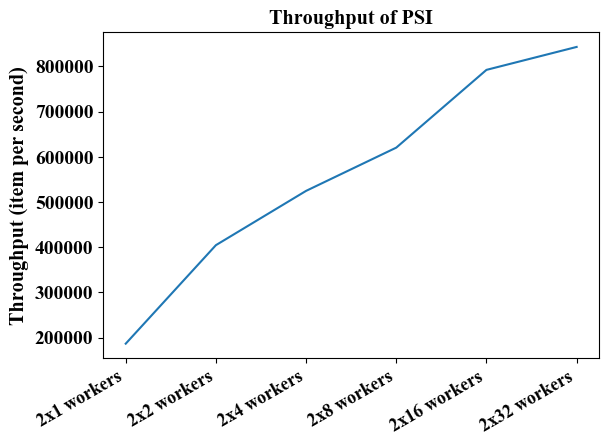}}
    \caption{The performance of Distributed PSI in the large-scale cluster environment.
    As the number of worker nodes increases, the execution time of the distributed PSI is reduced, while the throughput is improved}
    \label{performancePSI}
\end{figure}

We also observed the training time of distributed vertical FL decreases while worker nodes increases. Because of the inevitable communication cost, the throughput per worker of 1-worker-system is higher than a 32-worker-system. The training time dropped significantly when the number of workers increases from one to two (25865 seconds to 12056 second). With the workers continually increasing, the training time decreasing becomes inert. When the number of workers increases, the overall throughput of the distributed vertical FL is also augmented. However, because of the communication overhead, the increment becomes less significant when the number of workers becomes big enough.

\begin{table*}[ht]
    \centering
    \begin{tabular}{ |c|c|c|c|c|c| }
     \hline
     & workers per party & 1 worker & 2 worker & 4 worker & 8 worker \\
     \hline
     \multirow{3}{*}{Machine A} 
      & peak CPU spike (\%) & {\bf 98.91} & {\bf 98.59} & 97.09 & 93.48 \\ \cline{2-6}
      & memory usage (G)    & {\bf 61.95} & 51.3 & 48 & 45.37\\ \cline{2-6}
      & Throughput      & 1105 & 1505 & 1101 & 772\\ \hline
      
      \multirow{3}{*}{Machine B} 
      & peak CPU spike (\%) & 60.39 & 39.48 & 27.91 & 15.98\\ \cline{2-6}
      & memory usage (G)    & 18.78 & 16.82 & 15.83 & 14.82\\ \cline{2-6}
      & Throughput      & 809 & 324 & 248 & 175\\ \hline
      
      \multirow{3}{*}{Machine C} 
      & peak CPU spike (\%) & Null & {\bf 98.65} & 97.81 &  96.74 \\ \cline{2-6}
      & memory usage (G)    & Null & 23.29 & 18 & 15.14\\ \cline{2-6}
      & Throughput      & Null & 1843 & 1828 & 1921\\ \hline
      
      \multirow{3}{*}{Machine D} 
      & peak CPU spike (\%) & Null & 34.8 & 18.37 & 17.03\\ \cline{2-6}
      & memory usage (G)    & Null & 12.15 & 11.53 & 10.5\\ \cline{2-6}
      & Throughput      & Null & 400 & 262 & 165\\ \hline
    \end{tabular}
    \caption{Hardware usage of DVFL}
    \label{table:hardware}
\end{table*}

\begin{figure}[t]
    \centering
    \includegraphics[width=0.45\textwidth]{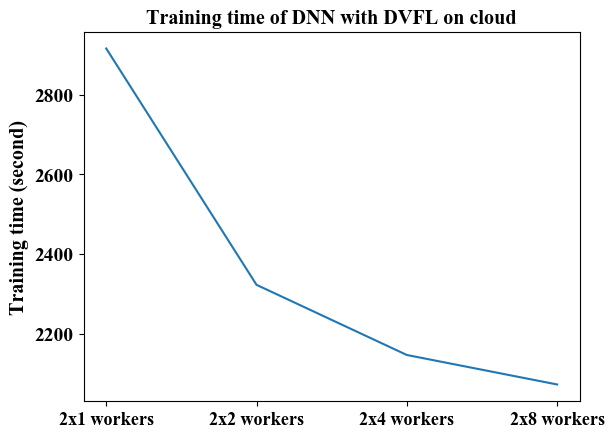}
    \caption{The training time of Deep Neural Networks (DNN) with DVFL and different numbers of workers on Baidu AI Cloud}
    \label{fig:nonSupercomputerRuntime}
\end{figure}

Additionally, we examined the execution time of the distributed PSI. In this experiment, we put $5 \times 10^8$ rows of data in Party A and $2 \times 10^7$ rows of data in Party P in the large-scale cluster environment. As shown in Figure \ref{performancePSI}, we obtain similar tendency in terms of the execution time and the throughput as that of DVFL. The execution time decreases from 2680 seconds to 593 seconds while the throughput increases from 186567 items per second to 843170 items per second.

We carry out another experiment to verify the performance of DVFL on the Baidu AI Cloud environment. In this experiment, Party A holds 4 $\times$ $10^8$ rows of data and Party P holds 2 $\times$ $10^7$ rows of data. The environment is given by four 32-core, 64 GB memory, and 160 GB storage machines in version 7.4 centOS Linux. Figure \ref{fig:nonSupercomputerRuntime} shows the execution time of 1, 2, 4, 8 workers per party in DVFL on above datasets and environment.

Besides drawing a similar conclusion as the previous experiment in the large-scale cluster, we also monitor the peak CPU spike, memory usage, and throughput (TPS) as shown in Table \ref{table:hardware}. In 1-worker per party case, Machine A handles Party A and Machine B handles Party P. In multi-worker cases, Machines A and B are responsible for Party A, while Machines C and D are responsible for Party P. We notice the CPU is overloaded when handling the magnitude of $10^8$ dataset in 1-worker system. Even the number of workers increases, the peak CPU spikes decrease lethargically. However, the memory usage is also close to full-filled to handle Party A in the 1-worker system (61.95 G out of 64 G). But the memory usage is effectively reduced while the number of workers is increased. The throughput drops as expected when the overall I/O remains, excepting few outliers. Hence, the limitation of CPU in a cloud environment could be a common bottleneck of DVFL when dealing with big data. 

\liu{In order to analyze the overhead of Paillier HE, we carry out an experimentation to compare the training time with vanilla vertical FL and that with Paillier HE. We set the number of rounds to 10, the learning rate to 0.05 and the batch size to 16. We vary the length of the public key in Paillier HE. As shown in Table \ref{table:publicKey}, the overhead of Paillier HE cannot be ignored. When we utilize the public key of 128 bits, the training is 8.9 times longer while the inference time is slightly longer (less than 1.4\%). When the length becomes 1024, the overhead can be much more significant (213 times longer). In practice, we exploit the public key of 128 bits to balance the security and the efficiency.}

\begin{table*}[ht]
    \centering
    \begin{tabular}{ |c|c|c|c| }
     \hline
     Type & Vanilla & HE (128) & HE (1024) \\
     \hline
      Training & 89 & 878 & 19021 \\ \hline
      Inference & 72 & 73 & 74 \\ \hline
    \end{tabular}
    \caption{Execution time for training and inference. The time unit is second. ``128'' and ``1024'' represent the length of public key.}
    \label{table:publicKey}
\end{table*}

\begin{figure}[t] 
    \centering
    {\includegraphics[width=0.45\textwidth]{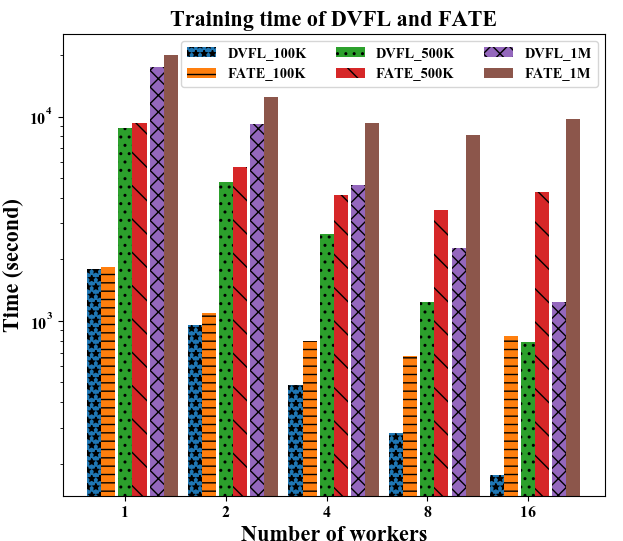}}
    \caption{
    The training time of DVFL and FATE with a single server at each party. }
    \label{fig:DVFLFATE}
\end{figure}

\begin{figure*}[ht] 
    \centering
    \begin{subfigure}{0.45\textwidth}
        \includegraphics[width=\textwidth]{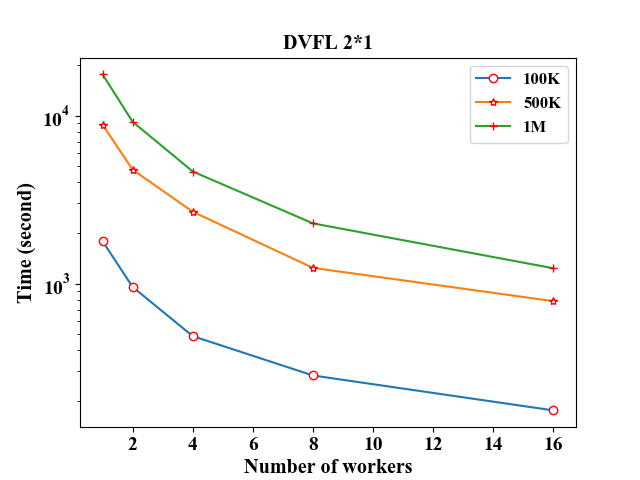}
            \caption{DVFL with 2*1 workers}
            \label{fig:DVFL21} 
    \end{subfigure}
    \begin{subfigure}{0.45\textwidth}
        \includegraphics[width=\textwidth]{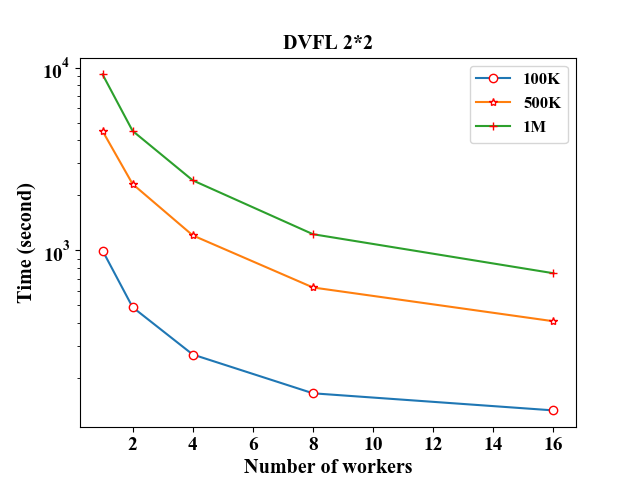}
            \caption{DVFL with 2*2 workers}
            \label{fig:DVFL22} 
    \end{subfigure}\\
    \begin{subfigure}{0.45\textwidth}
        \includegraphics[width=\textwidth]{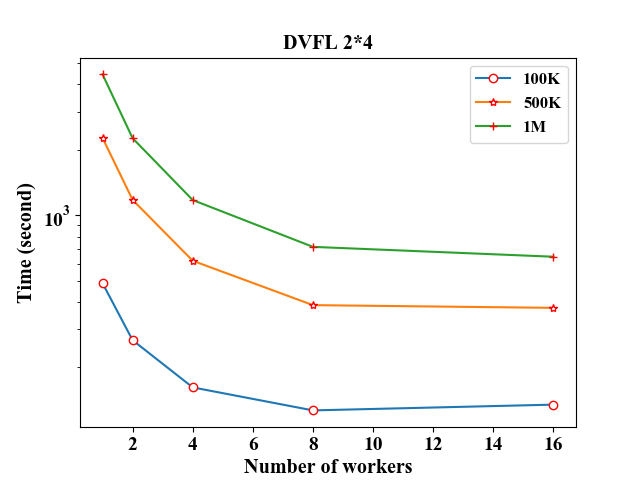}
            \caption{DVFL with 2*4 workers}
            \label{fig:DVFL24} 
    \end{subfigure}
    \begin{subfigure}{0.45\textwidth}
        \includegraphics[width=\textwidth]{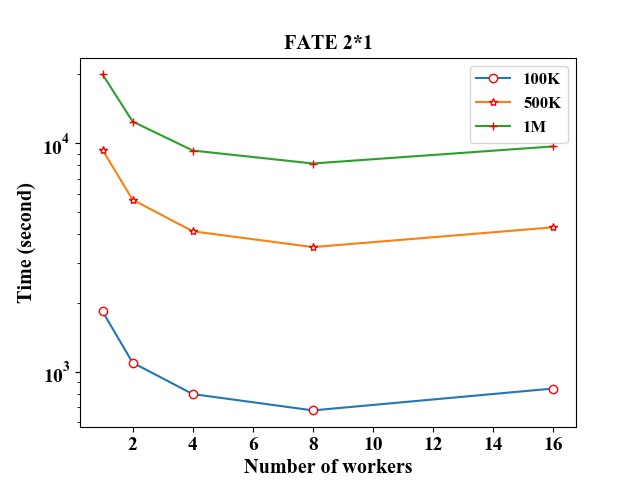}
            \caption{FATE with 2*1 workers}
            \label{fig:FATE21} 
    \end{subfigure}
    \caption{
    The execution time of DVFL and FATE with different numbers of workers.}
    \label{fig:executionTimeDVFLFATE}
\end{figure*}

\subsection{Comparison of DVFL and Baseline Frameworks}
\label{cfd}

In order to show the efficiency and scalability of DVFL, we compare the performance of DVFL with baseline frameworks, i.e., FATE and PyVertical. The major difference between DVFL and baseline frameworks is the implementation and the communication strategy. In DVFL, we exploit a peer to peer method. FATE utilizes a centralized strategy and only the parameter server can communicate with each other between the two parties. In addition, PyVertical does not provide support for multiple cores or workers for vertical federated learning.

The experiment is given by 128 GB memory, 16-core, 64-bit CentOS 8.2 Linux with x86\_64 architecture. To run DVFL and FATE, Party A and Party P each contains 50 thousand, 250 thousand and 500 thousand rows of data. To evaluate the performance, we record and compare the execution time of two parties running DVFL and FATE, while each party has 1, 2, 4 servers and each server has 1, 2, 4, 8, 16 virtual nodes respectively.

As shown in Figure \ref{fig:executionTimeDVFLFATE}, every time when the number of workers is doubled, the execution time of DVFL is reduced to nearly half of the original execution. Figure \ref{fig:DVFL21} shows the execution with one worker in each party with DVFL, Figure \ref{fig:DVFL22} shows the execution with two workers in each party with DVFL and Figure \ref{fig:DVFL24} shows the execution with four worker in each party with DVFL. When the number of workers increases, the overall execution time is significantly reduced. However, as shown in Figure \ref{fig:DVFL24}, the execution time decreases slowly when number of workers increases from 8 to 16, the execution time is not reduced due to the overhead of communication and synchronization.

Figure \ref{fig:FATE21} shows the execution time of FATE with a server in each party. When the number of workers increases, the execution time is reduced when the number of workers is smaller or equal to 8. When the number of workers increases from 8 to 16, the execution time is augmented as well. This is because the bottleneck of the centralized parameter servers that realize the communication between two parties. In addition, to the best of our knowledge, FATE cannot support the execution with multiple servers at each party. 

The comparison of DVFL and FATE is shown in Figure \ref{fig:DVFLFATE}, when the amount of data and the number of workers are the same, the execution time of FATE is significantly longer than that of DVFL (up to 6.8 times longer). In addition, the difference becomes more significant with more data to process. In addition, the execution of DVFL can be up to 11.6 times faster than the best performance of FATE with the consideration of multiple servers in each party.

\begin{figure}[t] 
    \centering
    {\includegraphics[width=0.45\textwidth]{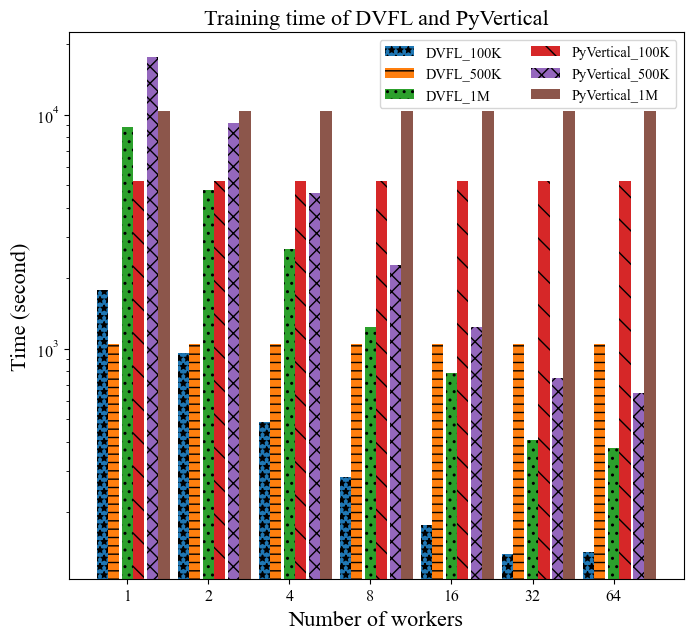}}
    \caption{
    The training time of DVFL and PyVertical with multiple cores or workers at each party.}
    \label{fig:DVFLPyVertical}
\end{figure}

\liu{As shown in Figure \ref{fig:DVFLPyVertical}, when the amount of data is the same, the execution of PyVertical \cite{romanini2021pyvertical} is shorter (up to 41.4\%) than DVFL. Please note that this difference is brought by the implementation of Paillier HE in DVFL while PyVertical only exploits DP. Paillier HE can well protect the security of the data while DP is much faster than Paillier HE in practice. However, when exploiting multiple cores at both party, DVFL can significantly outperform PyVertical (up to 15.1 times faster) when the data is partitioned and processed in parallel.}

\section{Conclusion and Perspective}
\label{sec:conclusion}
In this paper, we propose a Distributed Vertical Federated Learning (DVFL) approach, which exploit distributed architecture for the execution of PSI and vertical federated learning. We exploit a peer-to-peer communication strategy, which corresponds to efficient distributed execution.
We carried out extensive experiments to show the efficiency and scalability of DVFL with different scales of data, diverse numbers of servers in two environment, i.e., a large-scale cluster environment and a cloud environment.
We examined the performance with one party holding the order of $10^8$ and another holding the order of $10^7$ datasets in both the large-scale cluster and the cloud environments. We demonstrated the acceleration of multiple servers with DVFL against non-distributed deep VFL in two parties scenario. In addition, we monitored the hardware usage and evaluated the possible performance bottleneck of DVFL. Furthermore, we compare the execution time of DVFL with \liu{state-of-the-art frameworks}, i.e., FATE and PyVertical. The results shown that the advantage of DVFL can be up to 6.8 faster with single server at each party and \liu{15.1} times faster with multiple servers. 

In the future, we plan to extend DVFL to multiple parties and meanwhile to improve its performance. In addition, the interpretability \cite{li2021interpretable,liu2015survey} is also of much importance for federated learning, we intend to enable interpretability within DVFL in the future. 




\nocite{*}
\bibliography{references}

\begin{thebibliography}{10}
\providecommand \doibase [0]{http://dx.doi.org/}%

\bibitem{liu2022distributed}
Liu J, Huang J, Zhou Y, et al. From distributed machine learning to federated
  learning: a survey. {\it Knowledge and Information Systems} 2022\string;
  64(4)\string: 885-917.

\bibitem{yang2019federated}
Yang Q, Liu Y, Chen T, Tong Y. Federated machine learning: Concept and
  applications. {\it ACM Trans. on Intelligent Systems and Technology ({TIST})}
  2019\string; 10(2)\string: 1-19.

\bibitem{CCL}
{Standing Committee of the National People's Congress} . Cybersecurity Law of
  the People’s Republic of China.
  \url{https://www.newamerica.org/cybersecurity-initiative/digichina/blog/translation-cybersecurity-law-peoples-republic-china/};
  .
\newblock Online; accessed 22/02/2021.

\bibitem{GDPR}
{Official Journal of the European Union} . General data protection regulation.
  \url{https://eur-lex.europa.eu/legal-content/EN/TXT/PDF/?uri=CELEX:32016R0679};
  .
\newblock Online; accessed 12/02/2021.

\bibitem{chik2013singapore}
Chik WB. The Singapore Personal Data Protection Act and an assessment of future
  trends in data privacy reform. {\it Computer Law \& Security Review}
  2013\string; 29(5)\string: 554--575.

\bibitem{CCPA}
California Consumer Privacy Act Home Page. \url{https://www.caprivacy.org/}; .
\newblock Online; accessed 14/02/2021.

\bibitem{Gaff2014}
{Gaff} BM, {Sussman} HE, {Geetter} J. Privacy and Big Data. {\it Computer}
  2014\string; 47(6)\string: 7-9.

\bibitem{mcmahan2017communication}
McMahan B, Moore E, Ramage D, Hampson S, Arcas yBA. Communication-efficient
  learning of deep networks from decentralized data. In: Int. Conf. on
  Artificial Intelligence and Statistics ({AISTATS}). ; 2017\string: 1273-1282.

\bibitem{hardy2017private}
Hardy S, Henecka W, Ivey-Law H, et al. Private federated learning on vertically
  partitioned data via entity resolution and additively homomorphic encryption.
  {\it arXiv preprint arXiv:1711.10677} 2017.

\bibitem{nock2018entity}
Nock R, Hardy S, Henecka W, et al. Entity resolution and federated learning get
  a federated resolution. {\it arXiv preprint arXiv:1803.04035} 2018.

\bibitem{Liu2020FedVision}
Liu Y, Huang A, Luo Y, et al. Fedvision: An online visual object detection
  platform powered by federated learning. In: . 34. AAAI Conf. on Artificial
  Intelligence. ; 2020\string: 13172-13179.

\bibitem{Sabt2015TEE}
Sabt M, Achemlal M, Bouabdallah A. {Trusted Execution Environment: What It is,
  and What It is Not}. In: . 1. IEEE Int. Conf. on Trust, Security and Privacy
  in Computing and Communications. ; 2015\string: 57-64.

\bibitem{armknecht2015cryptoeprint}
Armknecht F, Boyd C, Carr C, et al. A Guide to Fully Homomorphic Encryption.
  Cryptology ePrint Archive, Report 2015/1192;  2015.
\newblock \url{https://eprint.iacr.org/2015/1192}.

\bibitem{li2021security}
Li B, Micciancio D. On the security of homomorphic encryption on approximate
  numbers. In: Annual Int. Conf. on the Theory and Applications of
  Cryptographic Techniques. ; 2021\string: 648--677.

\bibitem{zhang2020batchcrypt}
Zhang C, Li S, Xia J, Wang W, Yan F, Liu Y. BatchCrypt: Efficient Homomorphic
  Encryption for Cross-Silo Federated Learning. In: ; 2020\string: 493--506.

\bibitem{pinkas2018scalable}
Pinkas B, Schneider T, Zohner M. Scalable private set intersection based on OT
  extension. {\it ACM Trans. on Privacy and Security (TOPS)} 2018\string;
  21(2)\string: 1--35.

\bibitem{chen2017fast}
Chen H, Laine K, Rindal P. Fast private set intersection from homomorphic
  encryption. In: ACM SIGSAC Conf. on Computer and Communications Security. ;
  2017\string: 1243-1255.

\bibitem{gu2020federated}
Gu B, Dang Z, Li X, Huang H. Federated doubly stochastic kernel learning for
  vertically partitioned data. In: ACM SIGKDD Int. Conf. on Knowledge Discovery
  \& Data Mining. ; 2020\string: 2483--2493.

\bibitem{feng2020multi}
Feng S, Yu H. Multi-participant multi-class vertical federated learning. {\it
  arXiv preprint arXiv:2001.11154} 2020.

\bibitem{chen2020vafl}
Chen T, Jin X, Sun Y, Yin W. Vafl: a method of vertical asynchronous federated
  learning. {\it arXiv preprint arXiv:2007.06081} 2020.

\bibitem{guo2022microfedml}
Guo Y, Polychroniadou A, Shi E, Byrd D, Balch T. MicroFedML: Privacy Preserving
  Federated Learning for Small Weights. {\it Cryptology ePrint Archive} 2022.

\bibitem{2021Kairouz}
Kairouz P, McMahan HB, Avent B, et al. Advances and open problems in federated
  learning. {\it Foundations and Trends in Machine Learning} 2021\string;
  14(1-2)\string: 1-210.

\bibitem{FATE}
WeBank . Federated AI Technology Enabler ({FATE}).
  \url{https://github.com/FederatedAI/FATE}; .
\newblock Online; accessed 16/02/2021.

\bibitem{romanini2021pyvertical}
Romanini D, Hall AJ, Papadopoulos P, et al. Pyvertical: A vertical federated
  learning framework for multi-headed splitnn. {\it arXiv preprint
  arXiv:2104.00489} 2021.

\bibitem{liu2021data}
Liu J, Mo L, Yang S, et al. Data Placement for Multi-Tenant Data Federation on
  the Cloud. {\it IEEE Transactions on Cloud Computing} 2021.

\bibitem{de2019data}
De~Oliveira DC, Liu J, Pacitti E. Data-intensive workflow management: for
  clouds and data-intensive and scalable computing environments. {\it Synthesis
  Lectures on Data Management} 2019\string; 14(4)\string: 1--179.

\bibitem{liu2022large}
Liu J, Dong D, Wang X, et al. Large-scale knowledge distillation with elastic
  heterogeneous computing resources. {\it Concurrency and Computation: Practice
  and Experience} 2022\string: e7272.

\bibitem{ahmed2022hyper}
Ahmed U, Lin JCW, Srivastava G. Hyper-Graph Attention Based Federated Learning
  Method For Mental Health Detection. {\it IEEE Journal of Biomedical and
  Health Informatics} 2022.

\bibitem{ahmed20215g}
Ahmed U, Lin JCW, Srivastava G. 5G-Empowered Drone Networks in Federated and
  Deep Reinforcement Learning Environments. {\it IEEE Communications Standards
  Magazine} 2021\string; 5(4)\string: 55-61.

\bibitem{ahmed2021federated}
Ahmed U, Srivastava G, Lin JCW. A federated learning approach to frequent
  itemset mining in cyber-physical systems. {\it Journal of Network and Systems
  Management} 2021\string; 29(4)\string: 1--17.

\bibitem{gaurav2022security}
Gaurav A, Psannis K, Perakovi{\'c} D. Security of cloud-based medical internet
  of things (miots): a survey. {\it Int. Journal of Software Science and
  Computational Intelligence ({IJSSCI})} 2022\string; 14(1)\string: 1-16.

\bibitem{rajabi2019exposing}
Rajabi E, Greller W. Exposing social data as linked data in education. {\it
  Int. Journal on Semantic Web and Information Systems ({IJSWIS})} 2019\string;
  15(2)\string: 92-106.

\bibitem{ahmed2022reliable}
Ahmed U, Srivastava G, Lin JCW. Reliable customer analysis using federated
  learning and exploring deep-attention edge intelligence. {\it Future
  Generation Computer Systems} 2022\string; 127\string: 70--79.

\bibitem{sejdiu2020integration}
Sejdiu B, Ismaili F, Ahmedi L. Integration of semantics into sensor data for
  the IoT: a systematic literature review. {\it Int. Journal on Semantic Web
  and Information Systems ({IJSWIS})} 2020\string; 16(4)\string: 1-25.

\bibitem{wang2020cloud}
Wang Z, Yang Y, Liu Y, Liu X, Gupta BB, Ma J. Cloud-based federated boosting
  for mobile crowdsensing. {\it arXiv preprint arXiv:2005.05304} 2020.

\bibitem{liu2022Efficient}
Zhou C, Liu J, Jia J, et al. Efficient Device Scheduling with Multi-Job
  Federated Learning. {\it AAAI Conf. on Artificial Intelligence} 2022\string;
  36(9)\string: 9971-9979.

\bibitem{liu2022multi}
Liu J, Jia J, Ma B, et al. Multi-Job Intelligent Scheduling With Cross-Device
  Federated Learning. {\it IEEE Transactions on Parallel and Distributed
  Systems (TPDS)} 2022\string; 34(2)\string: 535-551.

\bibitem{Zhang2022FedDUAP}
Zhang H, Liu J, Jia J, Zhou Y, Dai H. {FedDUAP}: Federated Learning with
  Dynamic Update and Adaptive Pruning Using Shared Data on the Server. In: Int.
  Joint Conf. on Artificial Intelligence ({IJCAI}). ; 2022\string: 1-7.
\newblock To appear.

\bibitem{che2022federated}
Che T, Zhang Z, Zhou Y, et al. Federated Fingerprint Learning with
  Heterogeneous Architectures. In: IEEE Int. Conf. on Data Mining (ICDM). ;
  2022\string: 31--40.

\bibitem{jin2022accelerated}
Jin J, Ren J, Zhou Y, Lyu L, Liu J, Dou D. Accelerated federated learning with
  decoupled adaptive optimization. In: Int. Conf. on Machine Learning (ICML). ;
  2022\string: 10298--10322.

\bibitem{Li2022FedHiSyn}
Li G, Hu Y, Zhang M, et al. FedHiSyn: A Hierarchical Synchronous Federated
  Learning Framework for Resource and Data Heterogeneity. In: Int. Conf. on
  Parallel Processing ({ICPP}). ; 2022\string: 1-10.
\newblock To appear.

\bibitem{stergiou2021infemo}
Stergiou CL, Psannis KE, Gupta BB. InFeMo: flexible big data management through
  a federated cloud system. {\it ACM Transactions on Internet Technology
  ({TOIT})} 2021\string; 22(2)\string: 1-22.

\bibitem{zhang2018gelu}
Zhang Q, Wang C, Wu H, Xin C, Phuong TV. GELU-Net: A Globally Encrypted,
  Locally Unencrypted Deep Neural Network for Privacy-Preserved Learning.. In:
  Int. Joint Conf. on Artificial Intelligence ({IJCAI}). ; 2018\string:
  3933--3939.

\bibitem{dean2012large}
Dean J, Corrado GS, Monga R, et al. Large scale distributed deep networks. {\it
  Advances in Neural Information Processing Systems ({NeurIPS})} 2012\string:
  1232-1240.

\bibitem{low2014graphlab}
Low Y, Gonzalez JE, Kyrola A, Bickson D, Guestrin CE, Hellerstein J. Graphlab:
  A new framework for parallel machine learning. {\it arXiv preprint
  arXiv:1408.2041} 2014.

\bibitem{xing2015petuum}
Xing EP, Ho Q, Dai W, et al. Petuum: A new platform for distributed machine
  learning on big data. {\it IEEE Trans. on Big Data} 2015\string; 1(2)\string:
  49--67.

\bibitem{murray2013naiad}
Murray DG, McSherry F, Isaacs R, Isard M, Barham P, Abadi M. Naiad: a timely
  dataflow system. In: ACM Symposium on Operating Systems Principles ({SOSP}).
  ; 2013\string: 439--455.

\bibitem{kraska2013mlbase}
Kraska T, Talwalkar A, Duchi JC, Griffith R, Franklin MJ, Jordan MI. MLbase: A
  Distributed Machine-learning System. In: . 1. Conf. on Innovative Data
  Systems Research ({CIDR}). ; 2013\string: 2--1.

\bibitem{liu2021heterps}
Liu J, Wu Z, Yu D, et al. Heterps: Distributed deep learning with reinforcement
  learning based scheduling in heterogeneous environments. {\it arXiv preprint
  arXiv:2111.10635} 2021.

\bibitem{li2013parameter}
Li M, Zhou L, Yang Z, et al. Parameter server for distributed machine learning.
  In: . 6. Advances in Neural Information Processing Systems ({NeurIPS})
  Workshop. ; 2013\string: 2.

\bibitem{li2014communication}
Li M, Andersen DG, Smola AJ, Yu K. Communication efficient distributed machine
  learning with the parameter server. {\it Advances in Neural Information
  Processing Systems ({NeurIPS})} 2014\string; 27\string: 19--27.

\bibitem{freeman2004privatematching}
Freedman MJ, Nissim K, Pinkas B. Efficient Private Matching and Set
  Intersection. In:  Cachin C, Camenisch JL. \kern-2pt, eds. {\it Advances in
  Cryptology - EUROCRYPT}Advances in Cryptology - EUROCRYPT. ; 2004\string:
  1--19.

\bibitem{lu2020multi}
Lu L, Ding N. Multi-Party Private Set Intersection in Vertical Federated
  Learning. In: IEEE Int. Conf. on Trust, Security and Privacy in Computing and
  Communications ({TrustCom}). ; 2020\string: 707-714.

\bibitem{dong2013private}
Dong C, Chen L, Wen Z. When private set intersection meets big data: an
  efficient and scalable protocol. In: ACM SIGSAC Conf. on Computer \&
  communications security. ; 2013\string: 789--800.

\bibitem{gentry2009fully}
Gentry C. Fully homomorphic encryption using ideal lattices. In: Annual ACM
  symposium on Theory of computing. ; 2009\string: 169-178.

\bibitem{paillier1999public}
Paillier P. Public-key cryptosystems based on composite degree residuosity
  classes. In: Int. Conf. on the theory and applications of cryptographic
  techniques. ; 1999\string: 223--238.

\bibitem{broder2004network}
Broder A, Mitzenmacher M. Network applications of bloom filters: A survey. {\it
  Internet mathematics} 2004\string; 1(4)\string: 485--509.

\bibitem{naor2001efficient}
Naor M, Pinkas B. Efficient oblivious transfer protocols.. In: . 1. Symposium
  on Discrete Algorithms (SODA). ; 2001\string: 448--457.

\bibitem{ishai2008founding}
Ishai Y, Prabhakaran M, Sahai A. Founding cryptography on oblivious
  transfer-efficiently. In: Annual Int. cryptology Conf. ; 2008\string:
  572-591.

\bibitem{smola2010architecture}
Smola A, Narayanamurthy S. An architecture for parallel topic models. {\it Very
  Large Data Base ({VLDB}) Endowment} 2010\string; 3(1-2)\string: 703--710.

\bibitem{shasha1991optimizing}
Shasha D, Wang TL. Optimizing equijoin queries in distributed databases where
  relations are hash partitioned. {\it ACM Transactions on Database Systems
  (TODS)} 1991\string; 16(2)\string: 279-308.

\bibitem{zhao2019elastic}
Zhao X, Papagelis M, An A, Chen BX, Liu J, Hu Y. Elastic bulk synchronous
  parallel model for distributed deep learning. In: IEEE Int. Conf. on Data
  Mining ({ICDM}). ; 2019\string: 1504-1509.

\bibitem{GRPC}
Open Source High Performance Remote Procedure Call. \url{https://grpc.io/}; .
\newblock Online; accessed 7/05/2021.

\bibitem{kalita2014socket}
Kalita L. Socket programming. {\it Int. Journal of Computer Science and
  Information Technologies} 2014\string; 5(3)\string: 4802-4807.

\bibitem{chang2011libsvm}
Chang CC, Lin CJ. LIBSVM: a library for support vector machines. {\it ACM
  transactions on intelligent systems and technology (TIST)} 2011\string;
  2(3)\string: 1-27.

\bibitem{li2021interpretable}
Li X, Xiong H, Li X, et al. Interpretable Deep Learning: Interpretation,
  Interpretability, Trustworthiness, and Beyond. {\it arXiv preprint
  arXiv:2103.10689} 2021.

\bibitem{liu2015survey}
Liu J, Pacitti E, Valduriez P, Mattoso M. A survey of data-intensive scientific
  workflow management. {\it Journal of Grid Computing} 2015\string;
  13(4)\string: 457--493.

\end{thebibliography}

\end{document}